\documentclass[aps,preprint,onecolumn]{revtex4}
\usepackage{graphicx}

\begin{document}
\bibliographystyle{prsty}
\title{Optical chirality without optical activity: How surface plasmons give a twist to light}
\author{Aur\'{e}lien Drezet, Cyriaque Genet, Jean-Yves Laluet and Thomas W.~Ebbesen}
\affiliation{ ISIS, Laboratoire des Nanostructures, Universit\'e de Strasbourg and CNRS, 8 all\'{e}e Gaspard Monge, 67000 Strasbourg, France}

\date{\today}

\begin{abstract}
Light interacts differently with left and right handed three
dimensional chiral objects, like helices, and this leads to the
phenomenon known as optical activity. Here, by applying a
polarization tomography, we show experimentally, for the first
time in the visible domain, that chirality has a different optical
manifestation for twisted planar nanostructured metallic objects
acting as isolated chiral metaobjects. Our analysis demonstrate
how surface plasmons, which are lossy bidimensional
electromagnetic waves propagating on top of the structure, can
delocalize light information in the just precise way for giving
rise to this subtle effect.
\end{abstract}

\maketitle
\section{Introduction}
\indent Since the historical work of Arago~\cite{Arago} and
Pasteur~\cite{Pasteur}, chirality (the handedness of nature) has
generally been associated with optical activity, that is the
rotation of the plane of polarisation of light passing through a
medium lacking mirror symmetry~\cite{Hecht,Landau}. Optical
activity is nowadays a very powerful probes of structural
chirality in varieties of system. However, two-dimensional chiral
structures, such as planar molecules, were not expected to display
any chiral characteristics since simply turning the object around
leads to the opposite handedness (we remind that a planar
structure is chiral if it can not be brought into congruence with
its mirror image unless it is lifted from the plane). This
fundamental notion was recently challenged in a pioneering study
where it was shown that chirality has a distinct signature from
optical activity when electromagnetic waves interact with a 2D
chiral structure and that the handedness can be
recognized~\cite{Fedotov}. While the experimental demonstration
was achieved in the giga-Hertz (mm) range for extended 2D
structures, the question remained whether this could be achieved
in the optical range since the laws of optics are not simply
scalable when downsizing to the nanometer level. Here we report
genuine optical planar chirality for a single subwavelength hole
surrounded by left and right handed Archimedian spirals milled in
a metallic film.  Key to this finding is the involvement of
surface plasmons, lossy electromagnetic waves at the metal
surfaces, and the associated planar spatial
dispersion~\cite{Barnes,Genet}. Our results reveal how, in a
stringent and unusual way, this optical phenomenon connects
concepts of chirality, reciprocity and broken time symmetry.\\
\indent We remind that partly boosted by practical motivations,
such as the quest of negative refractive lenses~\cite{Pendry} or
the possibility to obtain giant optical activity for applications
in optoelectronics, there is currently a renewed
interest~\cite{Pendry,Papakostas,Schwanecke,Vallius,Gonokami,Canfield,Canfield2,Zhang,Decker,Plum,Rogacheva}
in the optical activity in artificial photonic media with planar
chiral structures. It was shown for instance that planar
gammadionic structures, which have by definition no axis of
reflection but a four-fold rotational
invariance~\cite{Papakostas,Vallius}, can generate optical
activity with giant gyrotropic
factors~\cite{Gonokami,Rogacheva,Plum,Decker}. Importantly, and in
contrast to the usual three dimensional (3D) chiral medium (like
quartz and its helicoidal structure~\cite{Hecht,Bose}), planar
chiral structures change their observed handedness when the
direction of light is reversed through the
system~\cite{Papakostas,Barron1}. This challenged Lorentz
principle of reciprocity~\cite{Landau} (which is known to hold for
any linear non magneto-optical media) and stirred up considerable
debate~\cite{Papakostas,Schwanecke,Gonokami,Barron2} which came to
the conclusion that optical activity cannot be a purely 2D effect
and always requires a small dissymmetry between the two sides of
the system~\cite{Gonokami,Rogacheva,Plum,Decker}. Nevertheless
Zheludev and colleagues did demonstrate in the GHz spectrum that a
pure 2D chiral structure lacking rotational symmetry can have an
optical signature which is distinct from optical
activity~\cite{Fedotov}. They went on to predict that it should be
possible to observe the same phenomena in the optical range by
scaling down their fish-scale structure and playing on localized
plasmons~\cite{Fedotov2}. Following a different strategy, we show
here that SP waves propagating on a 2D metal chiral grating
resonantly excited
by light provide an elegant solution to generate planar optical chirality in the visible.\\
\section{Experiments and Results}
\indent This is a challenging issue as it leads to two fundamental
points which are apparently incompatible. On the one hand, finding
such a 2D chiral effect in the optical domain is not equivalent to
a simple rescaling of the problem from the GHz to the visible part
of the spectrum. Indeed, losses in metal become predominant at the
nanometer scale so that the penetration length of light through
any chiral structure will become comparable to the thickness of
the structure. In-depth spatial dispersion along the propagation
direction of light will hence be induced, corresponding to the
usual 3D optical
activity~\cite{Vallius,Gonokami,Canfield,Canfield2,Zhang,Decker,Plum,Rogacheva}.
One thus expects optical activity, through the losses, to be a
more favorable channel than 2D optical chirality. On the other
hand, losses (i.e., broken time invariance at the macroscopic
scale) are necessary to guarantee planar chiral
behavior~\cite{Fedotov,Fedotov2}. With this in mind, we chose to
make single SP structures such as a single hole in an optically
thick metal film surrounded by an Archimedian spirals (Figure 1)
which can provide all the necessary ingredients for observing 2D
optical chirality. It is a 2D structure lacking point symmetry,
that is rotational and mirror invariances. At the same time, it
resonates due to coupling to surface plasmons which, as lossy
waves, represent a natural way for delocalizing information along
a planar interface, moving in-depth losses to the surface.
Importantly, the thickness of the metal film optically decouples
both interface~\cite{Degiron}, and consequently only the
structured chiral side is involved in the 2D optical chiral effect
reported here. Finally, the structures gives rise to enhanced
transmission~\cite{Genet} enabling high optical throughput for all
the characterization measurements.\\
\indent Using focus ion beam (FIB), we milled in an opaque gold
film a clockwise (right $\mathcal{R}$) or anticlockwise (left
$\mathcal{L}$) Archimedian spiral grooves around a central
subwavelength hole. The polar equation $(\rho,\theta)$ of the left
handed Archimedian spiral is $\rho=P\cdot\theta/(2\pi)$, and the
right handed enantiomeric spiral is obtained by reflection across
the y axis (see Fig.~1). \begin{figure}[h]
\centering\includegraphics[width=8.5cm]{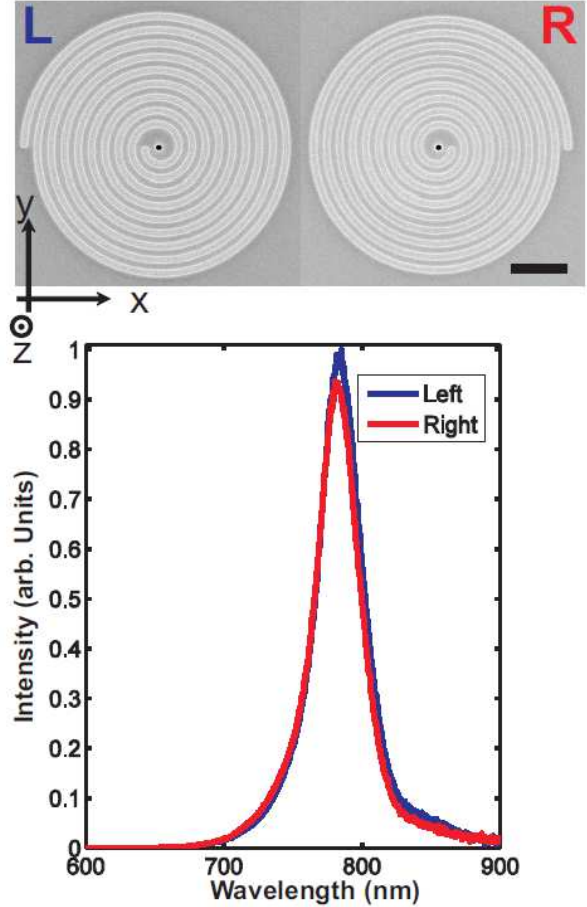}
\caption{Chiral plasmonic metamolecules. On the top panel:
scanning electron micrographs of the left (L) and right (R) handed
enantiomer (mirror image) planar chiral structures investigated.
The scale bare is 3 $\mu$m long. The parameters characterizing the
structure are the following: hole diameter $d=350$ nm, film
thickness $h=310$ nm, grating period $P=760$ nm, groove width
$w=370$ nm, and groove depth $s=80$ nm. The structures are milled,
with a focus ion beam, in a gold film deposited on a glass
substrate. On the bottom panel: transmission spectra at normal
incidence of individual left (blue curve) and right handed (red
curve) Archimede spirals illuminated from the air side. }
\end{figure}
The geometrical parameter $P$ is the radial grating period and we
take its value equal to the SP wavelength $\lambda_{SPP}\simeq
760$ nm (for an excitation at $\lambda\simeq 780$ nm). We recorded
optical transmission spectra at normal incidence with unpolarized
light for both isolated structures (Fig.~1). As it can be seen,
both enantiomers behave like resonant antennas with quasi
identical transmission properties. This resonant behaviour is a
direct indication of the SP excitation by the grating similarly to
what is observed
for circular antennas~\cite{Lezec}.\\
\indent To observe and fully characterize the optical signature of
planar chirality we perform a full polarization
tomography~\cite{Brehonnet,Altewischer} with the aim of
determining the 4$\times$4 Mueller matrix $\mathcal{M}$ associated
with each enantiomer. Experimental results
$\mathcal{M_L}^{\textrm{exp.}}$, and
$\mathcal{\mathcal{M_R}}^{\textrm{exp.}}$ respectively obtained
for left and right handed spirals are given in appendixes A and B.
Here the important point is that the degree of purity $F$ of the
Mueller matrices~\cite{Brehonnet} is near unity with
$F\left(\mathcal{M_L}^{\textrm{exp.}}\right)\simeq 0.967$ and
$F\left(\mathcal{M_R}^{\textrm{exp.}}\right)\simeq 0.939$. This
shows that the coherence in polarization is not degraded by the
structure and that we can therefore restrict our discussion to
Jones matrices~\cite{Brehonnet, Hecht}. In the convenient left
$|L\rangle$ and right $|R\rangle$ circular polarization basis,
these Jones matrices tie the excitation $
[E^{\textrm{in}}_L,E^{\textrm{in}}_R]$ to the transmitted  $
[E^{\textrm{out}}_L,E^{\textrm{out}}_R]$ electric fields. In the
case of planar chiral structures displaying 2D chiral activity,
they have the following form~\cite{Fedotov, Fedotov2}:
\begin{eqnarray}
\mathcal{J_L}^{\textrm{th.}}=\left(\begin{array}{cc} A & B
\\ C& A
\end{array}\right),\mathcal{J_R}^{\textrm{th.}}=\left(\begin{array}{cc} A &
C\\B & A
\end{array}\right),
\end{eqnarray}
where  $A$, $B$ and $C$ are complex valued numbers such that
$|B|\neq |C|$. This inequality account for chirality. Being non
diagonal, these matrices correspond to polarization converter
elements with no rotational invariance around the $z$ axis
(Fig.~1). They are thus fundamentally different from Jones
matrices associated with optical activity, e.g., gammadions.
Importantly the conditions $|B|\neq |C|$ implies the non unitarity
of $\mathcal{J}^{\textrm{th.}}_{\mathcal{L,R}}$ which means that
reversing the light path through the chiral structures is not
equivalent to reversing the time. From equation (1) we deduce the
associated theoretical forms for the Mueller matrices
$\mathcal{M_L}^{\textrm{th.}}$, $\mathcal{M_R}^{\textrm{th.}}$
(see appendix C) which are used to fit $\mathcal{J_L}$ and
$\mathcal{J_R}$ from experimental results. After normalization by
$A$ we deduce
\begin{eqnarray}
\mathcal{J_L}^{\textrm{fit}}=\left(\begin{array}{cc} 1.000 &
0.166+i0.221\\-0.131+i0.099 &1.000
\end{array}\right),\nonumber \\
\mathcal{J_R}^{\textrm{fit}}=\left(\begin{array}{cc} 1.000 &
-0.129+i0.098\\0.170+i0.230 &1.000
\end{array}\right).
\end{eqnarray}
These matrices indeed satisfy the chirality criteria of equation
(1) within the $\sim 1\%$ uncertainty evaluated from the degree of
purity of the Mueller matrice of the empty
setup.\\\begin{figure}[h]
\centering\includegraphics[width=14cm]{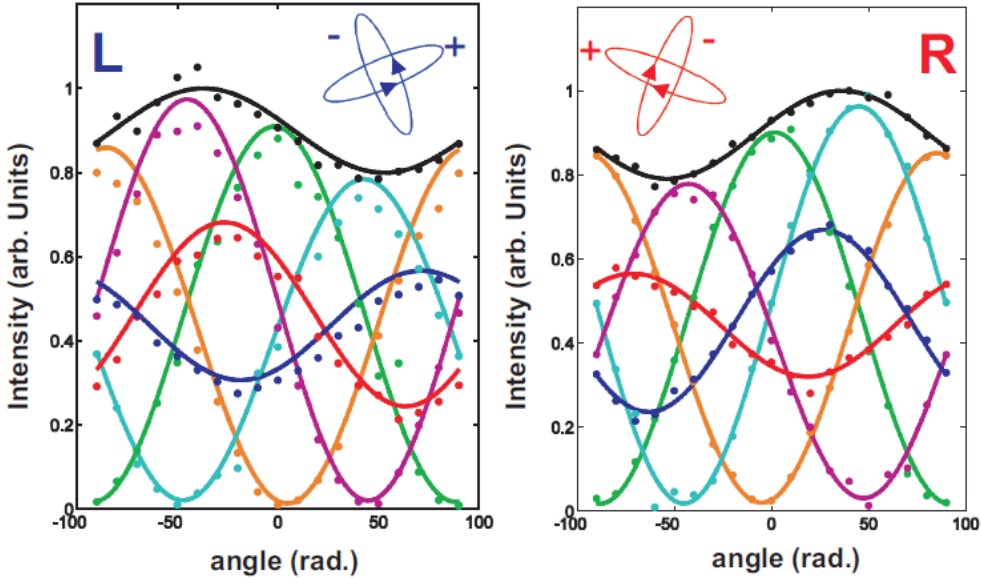}
\caption{Analysis of the polarization states for an input light
with variable linear polarization for both the left (left panel)
and right handed (right panel) individual chiral structures of
Fig.~1. The data points (acquired with a laser light at
$\lambda=780$ nm) are compared to the predictions from equation
(2) (continuous curves) for respectively the transmitted intensity
analyzed along the direction: $|x\rangle$ (green), $|y\rangle$
(yellow), $|+45^{\circ}\rangle$ (cyan), $|-45^{\circ}\rangle$
(magenta), $|L\rangle$ (red), and $|R\rangle$ (blue). The total
transmitted intensity is also shown (black). The symmetries
between both panel expected from group theory (see appendix D) are
observed experimentally. The insets show in each panel the
ellipses of polarization and the handedness (arrow) associated
with the two corotating eingenstates associated with the Jones
matrix $\mathcal{J_L}$ (blue) and $\mathcal{J_R}$ (red). }
\end{figure}
\section{Discussion and Conclusion}
\indent To illustrate the polarization conversion properties of
our chiral structures, we compare in Fig.~2 theory and experiment
when the input state is linearly polarized and when the output
transmitted intensity is analyzed along different orthogonal
directions. A good agreement between the measurements and the
theoretical predictions deduced from the Jones matrices (see
appendix D) is clearly seen, together with the mirror symmetries
between the two enantiomers. Importantly, these symmetries also
imply that for unpolarized light, and in complete consistency with
Fig.~1, the total intensity transmitted by the structures is
independent of the chosen enantiomer. Furthermore, the conversion
of polarization is well (geometrically) illustrated by using the
Poincar\'{e} sphere representation~\cite{Brehonnet}. Indeed, as
shown in Fig.~3, the Mueller matrix defines a geometrical
transformation which projects the unit Poincar\'{e} sphere, drawn
by the input Stokes vector, on an output closed surface with
typical radius $F\left(\mathcal{M}^{\textrm{exp.}}\right)\simeq 1$
in agreement with the absence of net depolarization as already
noticed (from theory, $F\left(\mathcal{M}^{\textrm{th.}}\right)=
1$ exactly). Data shown on Fig.~2 are also plotted on this sphere.
The input state draws a circle in the equator plane while the
output state (for each enantiomer) draws a circle in a different
plane, which center is not located at the center of the sphere.
This is a direct manifestation of planar chirality (see appendix
E). There is clearly an antisymmetrical behaviour between both
enantiomers.\begin{figure}[h]
\centering\includegraphics[width=8.5cm]{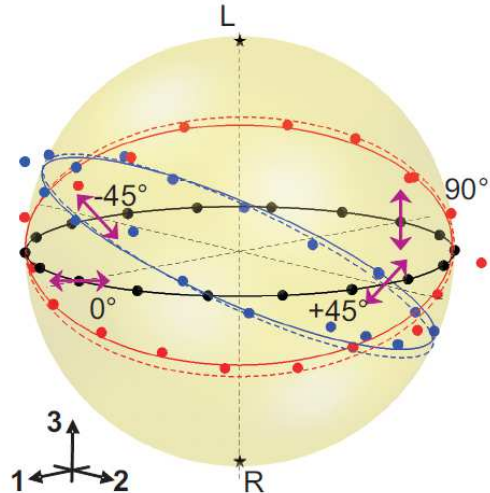}
\caption{Full polarization tomography. Poincar\'{e} sphere of unit
radius associated with the input state represented by the Stokes
vector $\mathbf{X}$~\cite{Hecht,Brehonnet}. Also shown are the
results of Fig.~2 for the left (blue) and right handed (red)
structures if the linearly polarized incident state draw the black
circle in the ($X_1$, $X_2$) equator plane of the input sphere.
Data points are compared with the predictions from
$\mathcal{M_{L,R}}^{\textrm{exp.}}$ (continuous curves) and of
equation (2) (dashed curves).}
\end{figure}
The good agreement between the experiment and the prediction of
equations (1,2) shows
the sensitivity of the polarization tomography method and the high reliability of the FIB fabrication.\\
\indent The degree of optical 2D chirality is quantified by
diagonalizing $\mathcal{J_L}^{\textrm{th.}}$ and
$\mathcal{J_R}^{\textrm{th.}}$. For
$\mathcal{J_L}^{\textrm{th.}}$, the eigenstates are
$|\pm_\mathcal{L}\rangle=\sqrt{B}|L \rangle \pm \sqrt{C}|R\rangle$
associated with the eigenvalues
$\lambda_\mathcal{L}(\pm)=A\pm\sqrt{(B\cdot C)}$. The eigenstates
for $\mathcal{J_R}^{\textrm{th.}}$ are obtained by permutation of
$B$ and $C$ with consequently
$\lambda_\mathcal{L}(\pm)=\lambda_\mathcal{R}(\pm)$. The scalar
product $\langle +_\mathcal{L}|
-_\mathcal{L}\rangle^{\textrm{th.}}=-\langle +_\mathcal{R}|
-_\mathcal{R}\rangle^{\textrm{th.}} =(|B|-|C|)/(|B|+|C|)$ is the
eigenstates Stokes parameter $S_3/S_0$ and provides a direct
measurement of the degree of optical chirality. It also evaluates
losses since the non-orthogonality of these two states is related
to the necessary non-unitarity of the Jones matrix for planar
chirality. We have $\langle +_\mathcal{L}|
-_\mathcal{L}\rangle^{\textrm{fit}}\simeq 0.255$ and $\langle
+_\mathcal{R}| -_\mathcal{R}\rangle^{\textrm{fit}}\simeq -0.277$,
which, within experimental uncertainties, are in good agrement
with the theoretical expectations. As shown in the insets of
Fig.~2, both eigenstates of each structure (e.g.,
$|\pm_{\mathcal{L,R}}\rangle$) can be represented by two ellipses
having the same axis ratio and the same handedness, but rotated
$90^{\circ}$ relative to each other. In agreement with the
theoretical predictions, these polarization ellipses for
$|\pm_{\mathcal{L}}\rangle$ and $|\pm_{\mathcal{R}}\rangle$ are
mirror reflections. This behaviour is significantly different from
the results obtained with optically active
media~\cite{Canfield,Canfield2,Gonokami,Rogacheva,Plum,Decker}
where the eigenstates associated with a given enantiomer have
opposite handedness~\cite{Landau}. This point, which reflects
itself in the symmetry property of chiral Jones matrices, namely
$\mathcal{J}_{\mathcal{L,R}}^{\textrm{th.}}(L,L)=\mathcal{J}_{\mathcal{L,R}}^{\textrm{th.}}(R,R)=A$,
has far reaching consequences, as pointed out in
reference~\cite{Fedotov}. It implies that a 2D plasmonic spiral
mimics a Faraday medium when we reverse the light path and this
even if the system, unlike a true Faraday medium, obeys rigorously
to the principle of reciprocity~\cite{Fedotov, Landau} (inversely,
one can show that equation (1) results from both this requirement
and the absence of mirror symmetry). It means that a photon coming
from the second side will probe a structure of opposite chirality.
After going through the structure and retracing back the light
path with a mirror normal to the axis, the polarization state will
be different at the end of journey from the initial one. This
would be impossible for an optically active medium and is solely
due to planar chirality. To summarize, our results therefore
demonstrate that 2D chirality is possible in the visible domain in
the absence of optical activity and add another element to the
promising plasmonic toolkit.
\section{Appendix A: Polarization tomography setup.}
We apply a procedure similar to the one considered in
\cite{Altewischer,Genet2} in order to record the Mueller matrix: a
collimated laser beam at $\lambda=785$ nm is focussed  normally on
the structure by using an objective $L_1$ ($\times 50$, numerical
aperture=0.55). The transmitted light is collected and
recollimated by using a second objective $L_2$ ($\times 40$,
numerical aperture=0.6). The input and output states of
polarization are respectively prepared and analyzed in the
collimated part of the light path by using polarizers, half wave
plates and quarter waveplates. A sketch of the setup is provided
below (see Fig.~4).\\
\begin{figure}[h]
\centering\includegraphics[width=12cm]{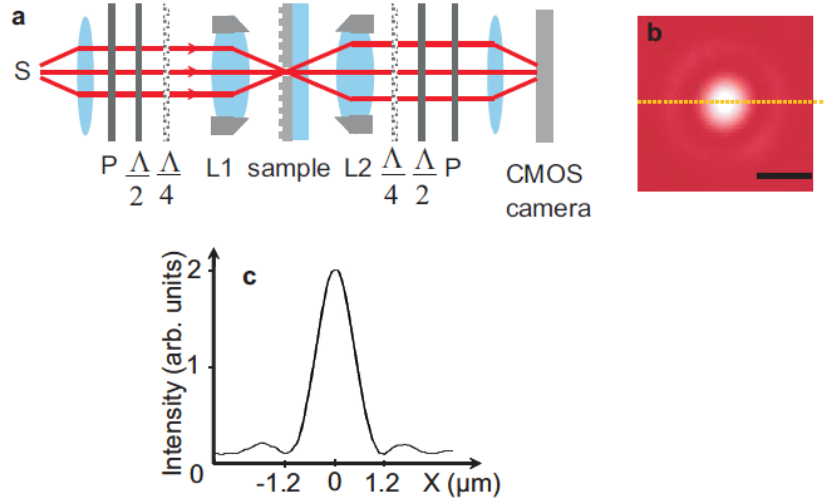}
\caption{Principle of the polarization experiment. (a), Sketch of
the optical set up described in the text. The images are recorded
by using a CMOS camera. (b), A typical image of the transmitting
nanohole showing the Airy spot associated with diffraction by the
optical microscope. The scale bar is  $2$ $\mu$m long. (c),
Crosscut of the intensity profile along the yellow dotted line
shown in (b). }
\end{figure}
The Mueller matrix is built by applying an experimental algorithm
equivalent to the one described in \cite{Brehonnet}. More
precisely, in order to write down the full Mueller matrix, we
measured here $6\times 6$ intensity projections corresponding to
the 6 unit vectors $|x\rangle$, $|y\rangle$,
$|+45^{\circ}\rangle$, $|-45^{\circ}\rangle$, $|L\rangle$, and
$|R\rangle$ for the input and the output polarizations. Actually
only 16 measures are needed to determine $\mathcal{M}$~
\cite{Brehonnet}. Our actual procedure is thus more than
sufficient to obtain $\mathcal{M}$.\\ The isotropy of the setup
was first checked by measuring the Mueller matrix
$\mathcal{M}^{\textrm{glass}}$ with a glass substrate. Up to a
normalization constant, we deduced that
$\mathcal{M}^{\textrm{glass}}$ is practically identical to the
identity matrix $\mathcal{I}$ with individuals elements deviating
by no more than 0.02. More precisely, the optical depolarization
(i.~e, the losses in polarization coherence) can be precisely
quantified through the degree of purity of the Mueller matrix
defined by \cite{Brehonnet}
$F\left(\mathcal{M}\right)=\left(\frac{\textrm{Tr}[\mathcal{M}^{\dagger}\mathcal{M}]-\mathcal{M}_{00}^{2}}{3\mathcal{M}_{00}^{2}}
\right)^{1/2}\leq 1$. Here we measured
$F\left(\mathcal{M}^{\textrm{glass}}\right)=0.9851$. It implies
that the light is not depolarized when going through the setup and
that consequently we can rely on our measurement
procedure for obtaining $\mathcal{M}$.\\
Two important points must be noted here: On the one hand  we
varied the incident illumination spot size on the sample between 2
and 20 $\mu$m without affecting the matrix, i.~e., without
introducing additional depolarisation. In the rest of the
experiment on chiral structures we consider the case of a large
gaussian spot with FWHM=20 $\mu$m in order to illuminate the whole
individual spiral. On the other hand, it can be observed that in
our experiments the polarization in the Airy spot (see Fig.~4b) is
homogeneous. This implies that we are actually doing the
polarization tomography of the central transmitting hole, i.~e.,
we are dealing only with the SU(2) point symmetry of the Mueller
Matrix. This situation clearly contrasts with previous SOP
tomography measurements on metallic hole arrays in which the
polarization degrees of freedom were mixed with spatial
information responsible for SPP-induced
depolarization~\cite{Altewischer}.
\section{Appendix B:
Experimental Mueller matrices.} The experimental Mueller matrices
deduced from the polarization tomography are after normalization
of $\mathcal{M}^{\textrm{exp.}}_{00}$:
\begin{eqnarray}
\mathcal{M_L}^{\textrm{exp.}}= \left(\begin{array}{cccc}
    1.000& 0.031 &  -0.107   &-0.029   \\
    0.029 &   0.958 &0.044    & -0.251  \\
    -0.105 & 0.037 &  0.953  & 0.287\\
  0.029  & 0.261  &  -0.282  &  0.809
\end{array}\right),\nonumber \\
\mathcal{M_R}^{\textrm{exp.}}= \left(\begin{array}{cccc}
   1.000& 0.035 &  0.111   &0.023   \\
    0.027 &   0.949 &-0.051    & 0.246  \\
    0.096 & -0.034 &  0.943  & 0.267\\
  -0.011  & -0.252  &  -0.277  &  0.745
\end{array}\right).
\end{eqnarray}
We have $F\left(\mathcal{M_L}^{\textrm{exp.}}\right)\simeq 0.967$
and $F\left(\mathcal{M_R}^{\textrm{exp.}}\right)\simeq 0.939$.\\
We must also note that the normalization used here neglects a
small additional coefficient of proportionality
$|\mathcal{M_L}^{\textrm{exp.}}_{ij}/\mathcal{M_R}^{\textrm{exp.}}_{ij}|\simeq
 0.954$ imputed to experimental errors and uncertainties.\\
We also recorded the Mueller matrix of the set up with the glass
substrate only. Up to a normalization factor we deduced
\begin{eqnarray} \mathcal{M}^{\textrm{glass}}=
\left(\begin{array}{cccc}
    \underline{1.0000} &   0.0060 &  -0.0040 &  -0.0070\\
   -0.0030 &   \underline{0.9851} &  -0.0010 &   0.0020\\
   -0.0020 &   0.0020 &   \underline{0.9965} &   0.0030\\
   -0.0050 &  -0.0040 &   0.0030 &   \underline{0.9821}\\
\end{array}\right)
\end{eqnarray}
which satisfies $\mathcal{M}^{\textrm{glass}}\simeq \mathcal{I}$
with $\mathcal{I}$ the identity matrix. It implies that the
optical set up do not induce depolarization and that consequently
we can rely on our measurement procedure for obtaining
$\mathcal{M}$.
\section{Appendix C: Theoretical Mueller matrices.}
The precise form of the theoretical Mueller matrice
$\mathcal{M_L}^{\textrm{th.}}$ deduced from equation (1) is
\begin{eqnarray}
\mathcal{M_L}^{\textrm{th.}}=\left(\begin{array}{cccc}
    \mathcal{M}^{\textrm{th.}}_{00}& \mathcal{M}^{\textrm{th.}}_{01}&  \mathcal{M}^{\textrm{th.}}_{02}  &\mathcal{M}^{\textrm{th.}}_{03}   \\
    \mathcal{M}^{\textrm{th.}}_{01}&   \mathcal{M}^{\textrm{th.}}_{11} &\mathcal{M}^{\textrm{th.}}_{12}    & \mathcal{M}^{\textrm{th.}}_{13} \\
    \mathcal{M}^{\textrm{th.}}_{02} & \mathcal{M}^{\textrm{th.}}_{12}&  \mathcal{M}^{\textrm{th.}}_{22} & \mathcal{M}^{\textrm{th.}}_{23}\\
  -\mathcal{M}^{\textrm{th.}}_{03}  & -\mathcal{M}^{\textrm{th.}}_{13}  &  -\mathcal{M}^{\textrm{th.}}_{23}  &   \mathcal{M}^{\textrm{th.}}_{33}
\end{array}\right).
\end{eqnarray}
with $\mathcal{M}^{\textrm{th.}}_{00}=(2|A|^2+|B|^2+|C|^2)/2$,
$\mathcal{M}^{\textrm{th.}}_{01}=Re[BA^{\ast}+AC^{\ast}]$,
$\mathcal{M}^{\textrm{th.}}_{02}=Im[AB^{\ast}+CA^{\ast}]$,
$\mathcal{M}^{\textrm{th.}}_{03}=(|C|^2-|B|^2)/2$,
$\mathcal{M}^{\textrm{th.}}_{11}=|A|^2+Re[B^{\ast}C]$,
$\mathcal{M}^{\textrm{th.}}_{12}=Im[B^{\ast}C]$,
$\mathcal{M}^{\textrm{th.}}_{13}=Re[AC^{\ast}-BA^{\ast}]$,
$\mathcal{M}^{\textrm{th.}}_{22}=|A|^2-Re[B^{\ast}C]$,
$\mathcal{M}^{\textrm{th.}}_{23}=Re[CA^{\ast}-AB^{\ast}]$,
$\mathcal{M}^{\textrm{th.}}_{33}=(2|A|^2-|B|^2-|C|^2)/2$. Similar
formula are obtained for $\mathcal{M_R}^{\textrm{th.}}$ after
permuting $B$ and $C$.\\
From the previous relations we deduce the useful equations (valid
for $\mathcal{M_L}^{\textrm{th.}}$)
\begin{eqnarray}
B/A=\frac{\mathcal{M}^{\textrm{th.}}_{01}-\mathcal{M}^{\textrm{th.}}_{13}}{\mathcal{M}^{\textrm{th.}}_{00}+\mathcal{M}^{\textrm{th.}}_{33}}+
i\frac{\mathcal{M}^{\textrm{th.}}_{23}-\mathcal{M}^{\textrm{th.}}_{02}}{\mathcal{M}^{\textrm{th.}}_{00}+\mathcal{M}^{\textrm{th.}}_{33}}\nonumber\\
C/A=\frac{M^{\textrm{th.}}_{01}+M^{\textrm{th.}}_{13}}{M^{\textrm{th.}}_{00}+M^{\textrm{th.}}_{33}}+
i\frac{\mathcal{M}^{\textrm{th.}}_{23}+\mathcal{M}^{\textrm{th.}}_
{02}}{\mathcal{M}^{\textrm{th.}}_{00}+\mathcal{M}^{\textrm{th.}}_{33}}.
\end{eqnarray} Together with equation (3) equation (6) allow us to fit
$B/A$ and $C/A$ if we replace $\mathcal{M}^{\textrm{th.}}$ by
$\mathcal{M_L}^{\textrm{exp.}}$ (a similar procedure is applicable
to $\mathcal{M_R}^{\textrm{exp.}}$  after permuting $B$ and
$C$).\\
The best fit we obtained (see equation (2)) are:
\begin{eqnarray} \mathcal{M_L}^{\textrm{fit}}=
\left(\begin{array}{cccc}
    1.000& 0.033 &  -0.116   &-0.023   \\
    0.033 &   0.951 &0.043    & -0.282  \\
    -0.116 & 0.043 &  0.951  & 0.304\\
  0.023  & 0.282  &  -0.304  &  0.902
\end{array}\right),\nonumber \\
\mathcal{M_R}^{\textrm{fit}}= \left(\begin{array}{cccc}
   1.000& 0.0359 &  0.125   &0.026   \\
    0.039 &   0.949 &-0.044    & 0.283  \\
    0.125 & -0.044 &  0.948  & 0.311\\
  -0.026  & -0.283  &  -0.311  &  0.897
\end{array}\right).
\end{eqnarray}
From theory we can deduce that
$F\left(\mathcal{M_{L,R}}^{\textrm{th.}}\right)=1$ (i.e., after
normalization by $\mathcal{M}^{\textrm{th.}}_{00})$. We have thus
$F\left(\mathcal{M_{L,R}}^{\textrm{fit}}\right)=1$
\section{Appendix D: Symmetries due to chirality [interpreting figure 2].}
Let $|\Psi_{\textrm{in}}\rangle=E_x|x\rangle+E_y|y\rangle$ and
$|\Psi_{\textrm{out}}\rangle=E'_x|x\rangle+E'_y|y\rangle$ be
respectively the incident and transmitted electric fields when we
consider the left handed planar chiral structure. We have
\begin{equation}
|\Psi_{\textrm{out}}\rangle=\hat{\mathcal{J_L}}|\Psi_{\textrm{in}}\rangle
\end{equation} where $\hat{\mathcal{J_L}}$ is the operator
associated with the Jones matrix $\mathcal{J_L}$. The mathematical
definition of planar chirality is that whatever the mirror
symmetry operation $\hat{\Pi}$ in the plane X-Y we have
$\hat{\mathcal{J}}\hat{\Pi}-\hat{\Pi}\hat{\mathcal{J}}\neq 0$. It
equivalently states that $\hat{\Pi}
\hat{\mathcal{J}}\hat{\Pi}^{-1}\neq \hat{\mathcal{J}}$. If we
consider for example the mirror reflection through the Y axis (see
Fig.~1) we have the matrix representation (in the cartesian basis)
${\Pi}={\Pi}^{-1}=\left(\begin{array}{cc} -1 &
 0
\\ 0& 1\end{array}\right)$ and consequently \begin{eqnarray}\hat{\Pi}
\hat{\mathcal{J_L}}\hat{\Pi}^{-1}= \hat{\mathcal{J_R}}\neq
\hat{\mathcal{J_L}}\end{eqnarray} which agrees with equation (1)
and constitutes an other optical definition of
chirality.\\
The previous equations are used in order to interpret the results
of Fig.~3 of the main article. Indeed from equations (8) and (9)
we obtain
\begin{equation}
\hat{\Pi}|\Psi_{\textrm{out}}\rangle=\hat{\mathcal{J_R}}\hat{\Pi}|\Psi_{\textrm{in}}\rangle.
\end{equation}
The input state considered in Fig.~2 is a linearly polarized light
$|\theta\rangle=\sin{(\theta)}|x\rangle+\cos{(\theta)}|y\rangle$
(the angle is measured relatively to the Y axis) and the
transmitted intensity projected along a direction of analysis
$|i\rangle$ (i.e, $|x\rangle$, $|y\rangle$, $|+45^{\circ}\rangle$,
$|-45^{\circ}\rangle$, $|L\rangle$, and $|R\rangle$) is written
$I_{i}^{(\textrm{Left})}(\theta)=|\langle
i|\Psi_{\textrm{out}}\rangle|^{2}=|\langle
i|\hat{\mathcal{J_L}}|\theta\rangle|^{2}$. Similarly we also write
$I_{i}^{(\textrm{Right})}(\theta)=|\langle
i|\hat{\mathcal{J_R}}|\theta\rangle|^{2}$. From equation (10) we
deduce:
\begin{equation}
\langle i'|\hat{\mathcal{J_L}}|\theta\rangle=\langle
i|\hat{\mathcal{J_R}}|-\theta\rangle,
\end{equation} where we used $|i'\rangle=\hat{\Pi}^{-1}|i \rangle=\hat{\Pi}|i \rangle$ and
$|-\theta\rangle=\hat{\Pi}|\theta \rangle$. We consequently have:
\begin{eqnarray}
I_{\textrm{total}}^{(\textrm{Left})}(\theta)=I_{\textrm{total}}^{(\textrm{Right})}(-\theta),\nonumber \\
I_{x,y}^{(\textrm{Left})}(\theta)=I_{x,y}^{(\textrm{Right})}(-\theta),\nonumber \\
I_{\pm 45^{\circ}}^{(\textrm{Left})}(\theta)=I_{\mp
45^{\circ}}^{(\textrm{Right})}(-\theta),\nonumber \\
I_{L,R}^{(\textrm{Left})}(\theta)=I_{R,L}^{(\textrm{Right})}(-\theta).
\end{eqnarray}
Such symmetries are clearly visible in Fig.~2 and correspond to a
direct signature of optical chirality in the planar systems
considered.
\section{Appendix E: Planar chirality on the Poincar\'{e} Sphere [interpreting figure 3]}
We remind that the Stokes parameters associated with a
polarization state of light $|\Psi\rangle$ are defined by
\begin{eqnarray}
S_0=I_x+I_y, &
S_1=I_x-I_y\nonumber\\
S_2=I_{+45^{\circ}}-I_{-45^{\circ}}, &
S_3=I_{\textrm{L}}-I_{\textrm{R}},
\end{eqnarray}
where $I_i$ are projection measurement along the direction $i$,
i.e, $I_{i}=|\langle i|\Psi\rangle|^{2}$. The Stokes vector
$\mathbf{X}$ is a convenient representation of such a state. We
have $\mathbf{X}=X_1\mathbf{x}_1+X_2\mathbf{x}_2+X_3\mathbf{x}_3$
with $X_1=S_1/S_0$, $X_2=S_2/S_0$, $X_3=S_3/S_0$ and with
($\mathbf{x}_1$, $\mathbf{x}_2$, $\mathbf{x}_3$) a cartesian
orthogonal and normalized vector basis.\\
The coherent input state satisfies the
normalization~\cite{Brehonnet} $|\mathbf{X}|=1$, that is the
vector draw a Poincar\'{e} sphere of unit radius in the space
$X_1,X_2,X_3$. The transmitted output state after interaction with
the left or right handed structure is defined by the relation
\begin{eqnarray}
\left(\begin{array}{c}
    S_{\mathcal{L,R};0}\\
    S_{\mathcal{L,R};1}\\
    S_{\mathcal{L,R};2} \\
  S_{\mathcal{L,R};3}
\end{array}\right)=\mathcal{M_{L,R}}\left(\begin{array}{c}
    S_0\\
    S_1 \\
    S_2 \\
  S_3
\end{array}\right).
\end{eqnarray}
The output state defines a Stokes vector
$\mathbf{X}_{\mathcal{L,R}}$ such that
$|\mathbf{X}_{\mathcal{L,R}}|\leq 1$. A typical value for
this radius is given by $F\left(\mathcal{M_{L,R}}\right)$.\\
If the input state is linearly polarized the input Stokes vector
is:
\begin{eqnarray}
\mathbf{X}^{\textrm{in}}(\theta)= \left(\begin{array}{c}
    \cos{(2\theta)} \\
    \sin{(2\theta)}\\
    0
\end{array}\right),
\end{eqnarray}
and draw a circle $(\sum_\textrm{in})$ along the equator contained
in the plane $X_1,X_2$ of the unit radius Poincar\'{e} sphere.
Using equation (14) the output Stokes vector is now a function of
$\theta$: $\mathbf{X}_{\mathcal{L,R}}(\theta)$ drawing a closed
curve $(\sum_\mathcal{L,R})$ (see Fig.~3) which is the image,
through the Mueller matrix transformation, of the equator circle
$(\sum_\textrm{in})$ above mentioned. Importantly, since the
Mueller matrix $\mathcal{M}$ given by equation (5) represents a
linear relation connecting $\mathbf{X}^{\textrm{in}}$ to
$\mathbf{X}^{\textrm{out}}$, we conclude that the image of the
incident polarization state contained in the equator plane
$X_1,X_2$ through $\mathcal{M}$ must also be contained in a plane
in the space $X_1,X_2,X_3$.\\To analyze this point more in details
we consider the normalized Vector product
\begin{eqnarray}
\mathbf{n}_{\mathcal{L,R}}=\frac{(\mathbf{X}_{\mathcal{L,R}}(0)-\mathbf{X}_{\mathcal{L,R}}(2\pi/3))\times(\mathbf{X}_{\mathcal{L,R}}(0)-\mathbf{X}_{\mathcal{L,R}}(\pi/2))}{|(\mathbf{X}_{\mathcal{L,R}}(0)-\mathbf{X}_{\mathcal{L,R}}(2\pi/3))\times(\mathbf{X}_{\mathcal{L,R}}(0)-\mathbf{X}_{\mathcal{L,R}}(\pi/2))|}
\end{eqnarray} and we write it
\begin{eqnarray}
\mathbf{n}_{\mathcal{L,R}}=\left(\begin{array}{c}
  U_{\mathcal{L,R}}\\
    V_{\mathcal{L,R}}\\
   W_{\mathcal{L,R}}
\end{array}\right),
\end{eqnarray}
with
$|U_{\mathcal{L,R}}|^2+|V_{\mathcal{L,R}}|^2+|W_{\mathcal{L,R}}|^2=1$.
It represents a typical normal to the closed curve
$(\sum_\mathcal{L,R})$. We have
\begin{eqnarray}
\mathbf{n}_{\mathcal{L}}=\left(\begin{array}{c}
  0.2845 \\
    -0.3065\\
   -0.9084
\end{array}\right), \mathbf{n}_{\mathcal{R}}=\left(\begin{array}{c}
  0.2861 \\
    0.3139\\
    0.95053
\end{array}\right).
\end{eqnarray}
Actually, if each curve $(\sum_\mathcal{L,R})$ is contained in a
(different) plane $P_{\mathcal{L,R}}$ we must have
\begin{equation} \mathbf{n}_{\mathcal{L,R}}\cdot
(\mathbf{X}_{\mathcal{L,R}}(\theta)-\mathbf{X}_{\mathcal{L,R}}(0))=0\end{equation}
for every $\theta$. This was indeed checked numerically up to a
precision of $10^{-11}$. It was also checked that
$|\mathbf{X}_{\mathcal{L,R}}(\theta)|=1$ up to the same precision.
This proves that each curve $(\sum_\mathcal{L,R})$ must be a
circle. The equations of the two planes $P_{\mathcal{L,R}}$ are
given by $\mathbf{n}_{\mathcal{L,R}}\cdot
(\mathbf{X}-\mathbf{X}_{\mathcal{L,R}}(0))=0$ where $\mathbf{X}$
is the Stokes vector associated with a running point belonging to
each plane. We write
\begin{eqnarray}
U_{\mathcal{L,R}}X_{1}+V_{\mathcal{L,R}}X_{2}+W_{\mathcal{L,R}}X_{3}+D_{\mathcal{L,R}}=0
\end{eqnarray}
with $D_{\mathcal{L}}=-0.0237$ and $D_{\mathcal{R}}=-0.0266$.
$|D_{\mathcal{L},R}|$ represents the distance separating the
center of the circle $(\sum_\mathcal{L,R})$ to the origin of the
poincar\'{e} sphere.  This proves that the planes are not going
through the center of the sphere. It was checked after lengthy
calculations that if $|B|=|C|$ in the Jones matrix (see equation
(1)) then $D=0$. This shows that the property
$|D_{\mathcal{L},R}|\neq 0$ is a characteristic of planar
chirality (i.e, the condition $|B|\neq|C|$). The radius of each
circle $(\sum_\mathcal{L,R})$ is given by
$r_{\mathcal{L,R}}=\sqrt{(1-D_{\mathcal{L},R}^2)}$ and we have
$r_{\mathcal{L}}=0.9997$ and $r_{\mathcal{L}}=0.9996$ which are
slightly smaller than $r=1$ in agreement with the fact that
$P_{\mathcal{L,R}}$ are not going through the center of the
sphere.

\end{document}